\documentstyle[aps,12pt,floats,epsfig]{revtex}
\tighten
\begin{document}

\preprint{}

\title{Introducing Time Dependence into the Static Maxwell Equations}

\author{Avraham Gal} 

\address{Racah Institute of Physics, The Hebrew University,\\ 
Jerusalem 91904, Israel}

\date{\today}
\maketitle

\begin{abstract}
Using to a minimum extent special relativity input, and relying on the
Lorentz-force expression for the force acting on a charged particle
in motion under the influence of electric ($\bf E$) and magnetic ($\bf B$)
fields, the Maxwell curl equations are shown to follow from the covariance 
of the two Maxwell divergence equations 
${\bf \nabla \cdot E} = 4 \pi \rho$ 
and ${\bf \nabla \cdot B} = 0$. 

\end{abstract}

\section{Introduction}
\label{sect1:introduction}

The Maxwell equations in vacuum provide the theoretical framework for
studying electromagnetism. Most of the undergraduate introductory-physics 
textbooks on Electricity and Magnetism (EM) start by introducing the basic
observations made, prior to Maxwell, regarding electric charges
and currents. The Maxwell equations are then `derived' by unifying
these observations into physical laws governing the space- and time
evolution of the electric field $\bf E$ and the magnetic field $\bf B$ 
(for a non exhaustive list of representative textbooks, 
see Refs. \cite{Pur85,Oha89,AF92,CS95,YF96,HRW01}). 
The equations which one usually finds easier to motivate than
others are the two divergence equations

\begin{equation}
	{\bf \nabla \cdot E} = 4 \pi \rho \ ,
	\label{eq:max1}
\end{equation}
and
\begin{equation}
	{\bf \nabla \cdot B} = 0 \ .
	\label{eq:max2}
\end{equation}
Eq. (\ref{eq:max1}) is the differential form of Gauss' law, relating
the electric field $\bf E$ to its charge sources, where $\rho$ is the
charge density. Eq. (\ref{eq:max2}), for the magnetic field $\bf B$,
expresses the non observation, and consequently the non existence of
magnetic charges. One then proceeds to explain how magnetism is
generated by currents rather than by charges, resulting in the
magnetic-field Maxwell curl equation

 \begin{equation}
 	{\bf \nabla \times B} = {{4 \pi}\over{c}} {\bf J} \ ,
 	\label{eq:max3}
 \end{equation}
 whereas the electric field is curl free:

 \begin{equation}
  	{\bf \nabla \times E} = 0 \ .
  	\label{eq:max4}
  \end{equation}
  The constant $c$ in Eq. (\ref{eq:max3}) is the speed of light in
  vacuum. Equations (\ref{eq:max3}) and (\ref{eq:max4}) apply to {\it
  static} situations, where any time variation of the underlying
  charge density $\rho$ and current density $\bf J$, and hence of 
  $\bf E$ and $\bf B$ respectively, is excluded.
  Allowing for time dependence, the above Maxwell curl 
  equations become

  \begin{equation}
  	{\bf \nabla \times B} = {{4 \pi}\over{c}} {\bf J} +
  	{{1}\over{c}} {{\partial {\bf E}}\over{\partial t}} \ ,
  	\label{eq:max5}
  \end{equation}
  \begin{equation}
  	{\bf \nabla \times E} = -{{1}\over{c}} {{\partial {\bf B}}
  	\over {\partial t}} \ .
  	\label{eq:max6}
  \end{equation}
  However, it turns out to be more problematic, and certainly less
  transparent, to motivate Eqs. (\ref{eq:max5},\ref{eq:max6}) by the
  appropriate phenomena, e.g. adding the displacement current for `deriving'
  Eq. (\ref{eq:max5}) and electromagnetic induction experiments for `deriving'
  Eq. (\ref{eq:max6}). In particular, Galili and collaborators 
  \cite{GK97,GL00,GK01} have 
  recently drawn attention to the subtleties involved in relating 
  Faraday's law to Maxwell's equation (\ref{eq:max6}). 
 
  The purpose of this note is to show that one 
  may {\it derive} Eqs. (\ref{eq:max5},\ref{eq:max6}) directly from
  Eqs. (\ref{eq:max1},\ref{eq:max2}) using a minimum input of
  EM (the Lorentz force) and special relativity physics 
  material that is within the scope of normal electromagnetism curriculum, 
  by requiring that Eqs. (\ref{eq:max1},\ref{eq:max2}), due to charge 
  conservation, hold in any inertial coordinate frame.
  Specifically, we will require (i) a knowledge of and
  familiarity with the Lorentz force acting on a charge $q$ moving at
  an instantaneous velocity $\bf v$; and (ii) 
  a knowledge of what a Lorentz four vector means, particularly for
  space-time $({\bf r}, ct)$ and for momentum-energy $({\bf p},
  E/c)$, but also for the current-charge density four vector 
  $({\bf J}, c \rho)$, in terms of a
  Lorentz boost from one inertial frame to any other inertial one. 
  Once the Maxwell equations (\ref{eq:max5},\ref{eq:max6}) are
  derived from (\ref{eq:max1},\ref{eq:max2}), the physics contents
  and phenomenology of these former equations can be easily and
  systematically unfolded within a regular undergraduate introductory 
  physics course.

  \section{Auxiliary material}
  \label{sect:aux}

  \subsection{Lorentz transformation}
  \label{subsect:Lortrans}

  A boost with velocity $\bf V$ is applied in the $x$ (longitudinal)
  direction, without affecting the $y, z$ transverse directions:

  \begin{equation}
  	x' = \gamma (x - \beta ct) \ , \quad ct' = \gamma (ct - \beta x)
  	\label{eq:lor1} \ ,
  \end{equation}
  where $\beta = V/c$, $\gamma = (1 - \beta^{2})^{-1/2}$. From
  (\ref{eq:lor1}) one shows that the derivatives transform as follows:

\begin{equation}
{\partial  \over {\partial x'}}=\gamma \left( {{\partial  \over
{\partial x}}+\beta {1 \over c}{\partial  \over {\partial t}}} \right),\quad \
{1 \over c}{\partial  \over {\partial t'}}=\gamma \left( {{1 \over c}{\partial
\over {\partial t}}+\beta {\partial  \over {\partial x}}} \right) \ .
	\label{eq:lor2}
\end{equation}
The Lorentz transformation rule (\ref{eq:lor1}) applies also to the
four momentum $({\bf p}, E/c)$ and to the four current
density $({\bf J}, c \rho)$. In particular, for the latter one:

\begin{equation}
	J'_{x} = \gamma \left( J_{x} - \beta c \rho \right) \ , \quad 
        \ c\rho' = \gamma \left(c \rho - \beta J_{x}\right) \ .	
	\label{eq:lor3}
\end{equation}
In order to motivate the (perhaps less familiar) transformation rule
(\ref{eq:lor3}), one recalls the continuity equation 

\begin{equation}
       {\bf \nabla \cdot J} + {{\partial \rho}\over{\partial t}} = 0 \ ,
       \label{eq:cont}
\end{equation}
which is usually taught within the context of charge conservation in 
EM courses \cite{Pur85} 
and which can be written in a covariant form $\partial _{\mu} J ^{\mu} = 0$, 
implying that $J^{\mu}$ ($\mu = 1,2,3,4$) is a four vector.

\subsection{Lorentz force}
\label{subsect:lorentzforce}

This is the basic dynamic law which governs the motion of charged
particles under the influence of electromagnetic fields:

\begin{equation}
	{\bf F} = q \left( {\bf E} + {{1}\over{c}} {\bf v \times B}
	\right) \ .
	\label{eq:lor}
\end{equation}
In many respects, Eq. (\ref{eq:lor}) may be regarded as the operational 
definition of the electromagnetic fields \cite{Pur85}.
Below, we will use this Lorentz-force expression in order to
determine the Lorentz transformation properties of $\bf E$ and $\bf B$. 
A caveat of the present approach, as generally is the case when
relying on Eq. (\ref{eq:lor}), is that the constant $c$ is the same 
speed of light constant as the one appearing in the Lorentz transformation 
(\ref{eq:lor1}).

\subsection{Transformation properties of $\bf E$ and $\bf B$}
\label{subsect:transform}

It is shown in the Appendix that the form of the Lorentz force, Eq.
(\ref{eq:lor}), leads to the following Lorentz transformation
properties of the electric field $\bf E$ and the magnetic field
$\bf B$:

\begin{equation}
E'_x=E_x\  ,\   E'_y=\gamma \left( {E_y-\beta B_z} \right)\  ,\
E'_z=\gamma \left( {E_z+\beta B_y} \right)  \ ,
	\label{eq:lor4}
\end{equation}

\begin{equation}
B'_x=B_x\  ,\   B'_y=\gamma \left( {B_y+\beta E_z} \right)\  ,\
B'_z=\gamma \left( {B_z-\beta E_y} \right) \ .
	\label{eq:lor5}
\end{equation}

\section{The Maxwell equations}
\label{sect:max}

Here we show, using the auxiliary material of Sect. \ref{sect:aux},
how the Maxwell equation (\ref{eq:max1}) leads to the Maxwell
equation (\ref{eq:max5}), and how the Maxwell equation 
(\ref{eq:max2}) leads to the Maxwell equation (\ref{eq:max6}). 
The key assumption is that the laws of electromagnetism expressed by 
Eqs. (\ref{eq:max1}) and (\ref{eq:max2}) 
(essentially charge conservation) have the same form in any inertial 
frame of reference. We start with Eq. (\ref{eq:max1}) in the primed 
coordinate system, 

\begin{equation}
	{\bf \nabla' \cdot E'} = 4 \pi \rho' \ ,
	\label{eq:max1'}
\end{equation}
and express the primed quantities in terms of the unprimed ones, 
using Eqs. (\ref{eq:lor2},\ref{eq:lor3},\ref{eq:lor4}). One obtains:

\begin{equation}
\gamma \left( {{\partial  \over {\partial x}}} +\beta {1 \over c}{\partial
\over {\partial t}} \right)E_x+{\partial  \over {\partial y}}\gamma
\left( {E_y-\beta B_z} \right)+{\partial  \over {\partial z}}\gamma
\left( {E_z+\beta B_y} \right)={{4\pi } \over c}\gamma \left( {c\rho -\beta J_x}
\right) \ .
	\label{eq:no1}
\end{equation}
Identifying the terms on the l.h.s. which combine to $\gamma {\bf
\nabla \cdot E}$, and on the r.h.s. the term $\gamma 4 \pi \rho$,
we make use of Eq. (\ref{eq:max1}) in the unprimed coordinate system
to cancel out these terms, resulting in

\begin{equation}
{1 \over c}{\partial  \over {\partial t}}E_x-\left( {{\partial  \over
{\partial y}}B_z-{\partial  \over {\partial z}}B_y} \right)= -{{4\pi }
\over c}J_x\  ,
	\label{eq:no2}
\end{equation}
namely
\begin{equation}
	({\bf \nabla \times B})_{x} = {{4 \pi}\over{c}}J_{x} +
	{{1}\over{c}} {{\partial}\over{\partial t}} E_{x} \ ,
	\label{eq:no3}
\end{equation}
which in view of the arbitrariness in choosing the boost $(x)$
direction leads to the curl equation (\ref{eq:max5})

\begin{equation}
	{\bf \nabla \times B} = {{4 \pi}\over{c}}{\bf J} +
	{{1}\over{c}} {{\partial}\over{\partial t}} {\bf E} \ .
	\label{eq:max5'}
\end{equation}
It is worth noting that applying the divergence operation to both sides 
of the Maxwell equation (\ref{eq:max5'}) which we herewith proved 
starting from the Maxwell equation (\ref{eq:max1}), and making use of 
the latter, one immediately recovers the continuity equation 
(\ref{eq:cont}) which provides the full (non static) content of 
the underlying assumption of charge conservation.

Similarly, starting with Eq. (\ref{eq:max2}) in the primed coordinate
system,

\begin{equation}
	{\bf \nabla' \cdot B'} = 0 \ ,
	\label{eq:max2'}
\end{equation}
and expressing the primed quantities in terms of the unprimed ones 
using Eqs. (\ref{eq:lor2},\ref{eq:lor5}), one obtains:

\begin{equation}
\gamma \left( {{\partial  \over {\partial x}}+\beta {1 \over c}{\partial
\over {\partial t}}} \right)B_x+{\partial  \over {\partial y}}\gamma
\left( {B_y+\beta E_z} \right)+{\partial  \over {\partial z}}\gamma
\left( {B_z-\beta E_y} \right)=0\  .
	\label{eq:no4}
\end{equation}
The terms on the l.h.s. combining to $\gamma {\bf \nabla \cdot B}$
give rise to zero by Eq. (\ref{eq:max2}), while the remaining terms
lead to

\begin{equation}
{1 \over c}{\partial  \over {\partial t}}B_x+\left( {{\partial  \over
{\partial y}}E_z-{\partial  \over {\partial z}}E_y} \right)=0\  ,
	\label{eq:no5}
\end{equation}
which upon permuting cyclically the $x, y, z$ directions produces the
curl equation (\ref{eq:max6}):

 \begin{equation}
 	{\bf \nabla \times E} = - {1 \over c} {{\partial}\over{\partial
 	t}} {\bf B} \ .
 	\label{eq:max6'}
 \end{equation}

 \section{Discussion and conclusion}
 \label{sect:concl}

 We have shown that accepting the Lorentz force, 
 plus a minimal amount of special
 relativity material, it becomes possible to derive each of the
 (vector) Maxwell curl equations (\ref{eq:max5},\ref{eq:max6}) from one
 of the (scalar) Maxwell divergence equations (\ref{eq:max1},\ref{eq:max2}),
 respectively. This suggests a more economical way of getting into the
 unified framework of the Maxwell equations than is done in most of the other
 approaches of teaching an introductory EM course. 
 In order to understand how come, on two separate
 occasions, {\it one} divergence equation gives rise to {\it three}
 equations which are combined into a curl equation, we
 note that the Maxwell equations (\ref{eq:max1}) and (\ref{eq:max5})
 can be lumped covariantly together into the form

 \begin{equation}
{{\partial F^{\alpha \beta }} \over {\partial x^\beta }}+{{4\pi }
\over c}J^\alpha =0\  ,
    	\label{eq:cov1}
    \end{equation}
    where $F^{\alpha \beta}$ is the $4 \times 4$ antisymmetric field
    tensor and the greek indices run over the four space-time
    components \cite{Jac75,LL59}. 
    $J^{\alpha}$ is the current density four vector, see 
    Eq. (\ref{eq:lor3}). If the form (\ref{eq:cov1}) is satisfied
    for {\it one} of these four-vector components $\alpha$, 
    by covariance it is also satisfied for the other ones, 
    since no direction in the four dimensional space has preference
    over the other ones. Eq. (\ref{eq:max1}) corresponds to
    the time-like component, whereas the vector Eq. (\ref{eq:max5})
    corresponds to the three space-like components. 
    A corollary of Eq. (\ref{eq:cov1}), due to the antisymmetry of 
    $F^{\alpha \beta}$, is the continuity equation (\ref{eq:cont}),
    in agreement with the discussion in Sect. III following 
    Eq. (\ref{eq:max5'}). 

    Similarly, the Maxwell equations (\ref{eq:max2}) and (\ref{eq:max6}) 
    can be lumped covariantly together into the form

    \begin{equation}
\varepsilon ^{\alpha \beta \gamma \delta }{{\partial F_{\gamma \delta }}
\over {\partial x^{\beta}}}=0\  ,
    	\label{eq:cov2}
    \end{equation}
    where $\varepsilon ^{\alpha \beta \gamma \delta}$ is the fully
    antisymmetric unit tensor in four dimensions \cite{Jac75,LL59}. 
    Eq. (\ref{eq:max2})
    corresponds to the time-like component, whereas the vector Eq.
    (\ref{eq:max6}) corresponds to the three space-like components.

    Obviously, if students were familiar with the antisymmetric field
    tensor $F^{\mu \nu}$, and once Eqs. (\ref{eq:max1},\ref{eq:max2}) 
    are shown to be equivalently written as the
    time-like components of Eqs. (\ref{eq:cov1},~\ref{eq:cov2})
    respectively, then it would have been straightforward to argue
    for the validity of Eqs. (\ref{eq:max5},\ref{eq:max6})
    respectively, using only the argument of Lorentz covariance.
    However, since most introductory EM courses do not introduce
    $F^{\mu \nu}$, the right thing to do in such a case would be to
    follow the explicit construction outlined in the present note 
    which emphasizes the frame independence of charge conservation.

\section*{Acknowledgments}

Stimulating discussions with, and useful advice from, Igal Galili 
and Issachar Unna are gratefully acknowledged.

\section*{Appendix: derivation of Eqs. (12, 13) for boosting
    $\bf E$ and $\bf B$}

    Let $K$ denote the laboratory coordinate frame, where a charged
    particle (charge $q$) moves with an instantaneous velocity $\bf
    v$ which defines the $x$ axis. The Lorentz force acting on it is
    given by

    \begin{equation}
    	{\bf F} = q \left({\bf E} + {{\bf v} \over c} {\bf \times B}
    	\right) \ ,
    	\label{eq:lor'}
    \end{equation}
    where $\bf E$ and $\bf B$ are the electric and magnetic fields.
    Define the instantaneous rest frame of the particle, where

    \begin{equation}
    	{\bf F'} = q {\bf E'} \ .
    	\label{eq:lor*}
    \end{equation}
    The frame $K'$ moves with velocity ${\bf V} = {\bf v}$ 
    with respect to $K$.
    We assume a knowledge of the force transformation law which is
    derived by making use how the momentum $\bf p$ and time $t$
    transform and recalling that ${\bf F} = d {\bf p} / dt$. Thus,
    $F'_{x} = F_{x}$, leading to

    \begin{equation}
    	E'_{x} = E_{x}+ \left({{\bf v} \over c} {\bf \times B} \right)_{x}
    	  = E_{x} \ ,
    	\label{eq:no6}
    \end{equation}
    since $\bf v$ is parallel to the $x$ direction. Similarly,
    $F'_{y} = \gamma F_{y}$, leading to

    \begin{equation}
    	E'_{y} = \gamma E_{y}+ \gamma \left({{\bf v} \over c} 
    {\bf \times B} \right)_{y} = \gamma \left(E_{y} - \beta B_{z}\right) \ ,
    	\label{eq:no7}
    \end{equation}
    and $F'_{z} = \gamma F_{z}$, leading to

    \begin{equation}
    	E'_{z} = \gamma E_{z}+ \gamma \left({{\bf v} \over c} {\bf \times B} 
        \right)_{z} = \gamma \left(E_{z} + \beta B_{y}\right) \ .
    	\label{eq:no8}
    \end{equation}
    Eqs. (\ref{eq:no6}-\ref{eq:no8}) give $\bf E'$ in terms of $\bf E$ 
    and $\bf B$, coinciding with Eq. (\ref{eq:lor4}) of the main
    text.

    In order to proceed with the derivation of Eq. (\ref{eq:lor5})
    for $\bf B'$ in terms of $\bf E$ and $\bf B$, we invert Eq.
    (\ref{eq:no7}) by noting that interchanging unprimed and primed
    quantities amounts to just changing the sign of $\beta$:

    \begin{equation}
    	E_{y} = \gamma (E'_{y} + \beta B'_{z}) \ .
    	\label{eq:no9}
    \end{equation}
    Substituting $E'_{y}$ from Eq. (\ref{eq:no7}), one gets

    \begin{equation}
    	E_{y} = \gamma \left[ \gamma (E_{y} - \beta B_{z}) + \beta
    	B'_{z} \right] \ ,
    	\label{eq:no10}
    \end{equation}
    and since $(1 - \gamma ^{2}) = - \beta^{2}\gamma^{2}$, the final
    outcome is

    \begin{equation}
    	B'_{z} = \gamma \left(B_{z} - \beta E_{y} \right) \ .
    	\label{eq:no11}
    \end{equation}
    Manipulating similarly on Eq. (\ref{eq:no8}), one gets

    \begin{equation}
    	B'_{y}=\gamma \left(B_{y} + \beta E_{z} \right) \ .
    	\label{eq:no12}
    \end{equation}
    Eqs. (\ref{eq:no11}-\ref{eq:no12}) for the transverse components 
    of $\bf B$ agree with the corresponding
    expressions in Eq. (\ref{eq:lor5}). Further arguments have to be
    invoked for showing that $B'_{x} = B_{x}$, thus completing the 
    derivation of Eq. (\ref{eq:lor5}). To spare a technically more 
    complicated procedure, we note that the transverse components of 
    $\bf B$ in Eq. (\ref{eq:lor5}) are obtained from those of $\bf E$ 
    in Eq. (\ref{eq:lor4}) by interchanging $\bf E$ and $\bf B$ plus 
    reversing the sign of $\beta$. Equation (\ref{eq:no6}) above for 
    the longitudinal component of $\bf E$, namely $E'_{x}=E_x$, then 
    suggests a similar rule $B'_{x}=B_x$ for the longitudinal component 
    of $\bf B$. It is useful to record at this stage, as follows from 
    Eqs. (\ref{eq:lor4},\ref{eq:lor5}), that 
    $\bf E \cdot B$ and $\bf {E}^2 - \bf {B}^2$ are invariant under 
    Lorentz transformations \cite{LL59}, leaving however the 
    demonstration of it to a later stage of the curriculum.

\end{document}